\documentclass[twocolumn,showpacs,preprintnumbers,amsmath,amssymb]{revtex4}
\usepackage{graphicx}% Include figure files
\usepackage{dcolumn}% Align table columns on decimal point
\usepackage{bm}% bold math

\begin{document}

%\preprint{APS/123-QED}

%\E-print{hep-th/070829?}

\title{Non-commutative phase space and its space-time symmetry}

\author{Kang Li}
\email{kangli@hznu.edu.cn}
\author{Sayipjamal Dulat}

 \email{dulat98@yahoo.com}

 \affiliation{Department of Physics,
Hangzhou Normal University, Hangzhou, 310036, China}

\affiliation{ School of Physics Science and Technology, Xinjiang
University, Urumqi, 830046, China \\
The Abdus Salam International Center for Theoretical Physics,
Trieste, Italy}

\date{\today}% It is always \today, today,

\begin{abstract}
First a description of 2+1 dimensional non-commutative(NC) phase
space is presented, where the deformation of the planck constant
is given. We find that in this new formulation, generalized Bopp's
shift has a symmetric representation and one can easily and
straightforwardly define the star product on NC phase space. Then
we define non-commutative Lorentz transformations both on NC space
and NC phase space. We also discuss the Poincare symmetry. Finally
we point out that our NC phase space formulation and the NC
Lorentz transformations  can be applicable to any even dimensional
NC space and NC phase space.


\end{abstract}

\maketitle

\section{ Introduction}
In recent years, there has been an increasing interest in the
study of physics on non-commutative space, because the effects of
the space non-commutativity may become significant in the extreme
situation such as at the string scale or at the TeV and even
higher energy level. There are many papers devoted to the study of
various aspects of the quantum field theory and quantum mechanics
on NC space, where space-space is non-commuting, but the
momentum-momentum is commuting, or on NC phase space, where both
space-space and momentum-momentum are non-commuting. For
references see \cite{1}-\cite{5}. Although quantum theories on  NC
space and NC phase space have been extensively studied in the
literature, the description of NC phase space is far from
complete, for example, it is not easy to define the star product
on NC phase space in the formulations of Refs.\cite{7} and
\cite{8}. Another important issue we want to discuss is the
space-time symmetries of the NC space and NC phase space. As we
know, the Lorentz symmetry plays a central role in any realistic
quantum field theory. There are different approaches in the
formulations of Lorentz and Poincare symmetries on NC space, for
references see \cite{6}-\cite{luki}. Ref.\cite{6} studied Lorentz
transformation on NC space and claimed that the NC gauge theories
are invariant under the NC Lorentz transformations. Because of the
singularity of matrix $\theta_{ij}$ the NC Lorentz transformation
in Ref.\cite{6} may not be applicable for 3+1 dimensional NC
space. Also there is no discussions about the Lorentz
transformation on NC phase space in the literature.

In this paper, we will give a description of NC phase space on
$2+1$ dimensions, where the star product can be easily defined. On
$2+1$ dimensional space-time, we extend the results in
Ref.\cite{6} to the NC phase space, and  we find that the new
formulation is  applicable to any even  space dimensions.

This paper is organized as follows: in Section 2, we present a
description of NC phase space on $2+1$ dimensions.
 In Section 3, we discuss NC
Lorentz and Poincare transformations both on NC space and NC phase
space. Conclusion remarks are given in the last section.

\section{The description of NC phase space}
 On NC space, the NC algebra is
\begin{eqnarray}\label{eq1}
[\hat{x}_\mu,\hat{x}_\nu]=i\theta_{\mu\nu} \;,~~~
[\hat{x}_\mu,\hat{p}_\nu]=i\hbar\delta_{\mu\nu}\;,~~~[\hat{p}_\mu,\hat{p}_\nu]=0
\;.
\end{eqnarray}
where the Greek indices $\mu$ and $\nu$ run from $0$ to $2$. In
order not to spoil unitarity \cite{gomis, alvarez} and
casuality\cite{seiberg}, one sets $\theta^{0\mu}=0$; in two
dimensions the constant anti-symmetric matrix element
$\theta_{ij}$ can be written as $\theta_{ij}=\epsilon_{ij}\theta
$,  and $\theta$ related to the space-space non-commutativity.

 Non-commutative field theories are constructed from
commutative field theories by replacing, in the action, the usual
multiplication product of fields with the star product of fields.
The star product between two fields is defined as
\begin{eqnarray}\label{eq2}
(f  \ast g)(x) &=&f(x) e^{  \frac{i}{2}
  \overleftarrow{\partial_\mu} \theta^{\mu\nu}\overrightarrow{\partial_\nu}
 } g(x) \nonumber\\ &=& f(x)g(x)
 + \frac{i}{2}\theta^{\mu\nu} \partial_\mu f \partial_\nu
 g+{\mathcal{ O}}(\theta^{2}) \;.
\end{eqnarray}
This star product can be replaced by a shift which is called
Bopp's shift£¬
\begin{equation}\label{eq3}
 \hat{x}_{\mu}=  x_{\mu}-\frac{1}{2}\theta_{\mu\nu}p^{\nu} \;,
 ~~~~~~~~~~\hat{p}_\mu=p_\mu \;.
\end{equation}
where  $x_\mu$ and $p_\nu$ are coordinates and momenta on
commuting space-time. After applying this shift, the effect caused
by space-space non-commutativity can be calculated in the
commuting space.

 The Bose-Einstein statistics on non-commutative quantum mechanics
 requires both space-space and  momentum-momentum non-commutativity
 \cite{8}. In the following we present our formulation to NC phase space.
 On NC phase space, we set the NC algebra as
\begin{eqnarray}\label{eq4}
[\hat{x}_\mu,\hat{x}_\nu]=i\theta_{\mu\nu} \;,~~
[\hat{x}_\mu,\hat{p}_\mu]=i\tilde{\hbar}\delta_{\mu\nu}
\;,~~[\hat{p}_\mu,\hat{p}_\nu]=i\bar{\theta}_{\mu\nu} \;.
\end{eqnarray}
On NC phase space we set also $\bar{\theta}^{0\mu}=0$,
$\bar{\theta}_{\mu\nu}$ is a very small constant anti-symmetric
matrix element, it reflects the non-commutativity of the momenta,
 and $\bar{\theta}_{ij}=\epsilon_{ij}\bar{\theta}$.
 The $\tilde{\hbar}$ in
Eq.(\ref{eq4}) is the deformation of Planck constant on NC phase
space, it has the form
\begin{equation}\label{eq5}
\tilde{\hbar}=\hbar+\frac{\theta\bar{\theta}}{4\hbar},
\end{equation}
and hence the momentum operator on NC phase space can be written
as
\begin{equation}\label{eq6}
\hat{p}_\mu=-i\tilde{\hbar}\frac{\partial}{\partial x^\mu}.
\end{equation}
From the above relations, we can obtain a generalized Bopp's shift
as
\begin{eqnarray}\label{eq7}
\hat{x}_\mu=x_\mu-\frac{1}{2\hbar}\theta_{\mu\nu}p^\nu,\\
\hat{p}_\mu=p_\mu+\frac{1}{2\hbar}\bar{\theta}_{\mu\nu}x^\nu.\label{eq7-1}
\end{eqnarray}
It is easy to check that the above generalized Bopp's shift is
consistent with the algebraic relation (\ref{eq4}). Then the star
product on NC phase space can be easily defined as
\begin{eqnarray}\label{eq8}
(f  \ast g)(x) &=& f(x,p) e^{  \frac{i}{2}
  \overleftarrow{\partial^x_\mu} \theta^{\mu\nu}\overrightarrow{\partial^x_\nu}
  + \frac{i}{2}
  \overleftarrow{\partial^p_\mu} \bar{\theta}^{\mu\nu}\overrightarrow{\partial^p_\nu}
 } g(x,p) \nonumber\\
 & = & f(x,p)g(x,p)
 + \frac{i}{2}\theta^{\mu\nu} \partial^x_\mu f(x,p) \partial^x_\nu
 g(x,p) \nonumber\\
 & + & \frac{i}{2}\bar{\theta}^{\mu\nu} \partial^p_\mu f(x,p) \partial^p_\nu
 g(x,p)+{\mathcal{ O}}(\theta^{2}).
\end{eqnarray}
In NC quantum mechanics and NC quantum field theory, the star
product between two fields on NC phase space can be replaced by
the generalized Bopp's shift (\ref{eq7}) for coordinates and
(\ref{eq7-1}) for momenta .

Now we would like to stress that our formulation above is well
defined on 2-dimensional NC phase space, and this formulation can
be generalized only to any even dimensional case.

\section{Lorentz transformation on non-commutative phase space}

Let's first discuss the Lorentz transformation on NC
space-time\cite{6}. In an analogous way as in commutative
space-time, on NC space one could introduce a NC Lorentz
transformation as $\hat{x}_\mu=\Lambda_\mu^\nu\hat{x}_\nu$. But
this kind of definition of Lorentz transformation is not
consistent with the algebra (\ref{eq1}), since it would require
$\theta_{\mu\nu} $ transforms as
$\Lambda_\mu^\alpha\Lambda_\nu^\beta\delta_{\alpha\beta}$, this
  makes little sense, because  $\theta_{\mu\nu}$ is a
constant and does not change under Lorentz transformation.

From the Bopp's shift (\ref{eq3}) on NC space one finds that
\begin{equation}\label{eq9}
x_\mu=\hat{x}_\mu+\frac{1}{2\hbar}\theta_{\mu\nu}\hat{p }^\nu \; ,
~~~~
 p_\mu=\hat{p}_\mu,
\end{equation}
 On commuting space-time  we can define a
Lorentz transformation as follows
\begin{equation}\label{eq10}
 x'_\mu=\Lambda_\mu^\nu x_\nu,
\end{equation}
which leaves the interval
\begin{equation}\label{s^2}
 s^2=\eta_{\mu\nu}x^\mu x^\nu
 \end{equation}
 invariant if
$\eta_{\mu\nu}\Lambda^\mu_\alpha\Lambda^\nu_\beta=\eta_{\alpha\beta}$.
Under the Lorentz transformation (\ref{eq10}), the momentum
$p_\mu$ transforms as a Lorentz vector
\begin{equation}\label{eq11}
 p'_\mu=\Lambda_\mu^\nu p_\nu \;.
\end{equation}
From the Bopp's shift (\ref{eq3}), we can obtain the following
Lorentz transformation on NC space-time which is induced by the
Lorentz transformation (\ref{eq10}) and (\ref{eq11})
\begin{eqnarray}\label{eq12}
\hat{x}'_\mu=
x'_\mu-\frac{1}{2\hbar}\theta_{\mu\nu}p'^\nu=\Lambda_\mu^\nu x_\nu
-\frac{1}{2\hbar}\theta_{\mu\nu}\Lambda^\nu_\rho p^\rho \nonumber \\
  =\Lambda_\mu^\nu \hat{x}_\nu +\frac{1}{2\hbar}\Lambda_\mu^\nu\theta_{\nu\rho}\hat{p}_\rho-\frac{1}{2\hbar}\theta_{\mu\nu}\Lambda^\nu_\rho
 {\hat p}^\rho.
\end{eqnarray}
The above equation  defines the non-commutative Lorentz
transformation on NC space-time.  From this transformation one may
note that rather than $\theta^{\mu\nu}$ transforms as a Lorentz
tensor, the $\theta^{\mu\nu}p_\nu$ transforms as a Lorentz vector.
So it is easy to check that the commutation relation (\ref{eq1})
on NC space-time is invariant under this transformation. And,
obviously, when $\theta^{\mu\nu}\rightarrow 0$, the NC Lorentz
transformation above becomes usual Lorentz transformation on
commuting space-time. From the shift (\ref{eq9}) and the Lorentz
invariant interval (\ref{s^2}), one finds the square of the NC
length
\begin{equation}\label{eq13}
s^2_{ncs}=\hat{x}^\mu \hat{x}_\mu
+\frac{1}{\hbar}\theta_{\mu\nu}\hat{x}^\mu \hat{p}^\nu
+\frac{1}{4\hbar^2}\theta^{\mu\alpha}\theta_{\mu\beta}\hat{p}_\alpha
\hat{p}^\beta.
\end{equation}
Straight forward calculation shows that the $s^2_{nc}$ is
invariant under NC Lorentz transformation (\ref{eq12}). So we have
defined a non-commutative Lorentz transformation which leaves
$s^2_{nc}$ invariant.

Now we are in  position to generalize the Lorentz transformation
on NC space to NC phase space.  From the generalized Bopp's shift
(\ref{eq7}) and (\ref{eq7-1}) on NC phase space, we obtain it's
inverse transformations
\begin{eqnarray}
x_\mu=\gamma (\hat{x}_\mu+\frac{1}{2\hbar}\theta_{\mu\nu}\hat{p}^\nu),\label{eq14}\\
p_\mu=\gamma
(\hat{p}_\mu-\frac{1}{2\hbar}\bar{\theta}_{\mu\nu}\hat{x}^\nu),\label{eq15}
\end{eqnarray}
where $\gamma =4\hbar^2/(4\hbar^2-\theta\bar{\theta})$. The
Lorentz transformations (\ref{eq10}) and (\ref{eq11}) induce the
following transformations on NC phase space
\begin{eqnarray}
\hat{x}'_\mu= \Lambda_\mu^\nu x_\nu
-\frac{1}{2\hbar}\theta_{\mu\nu}\Lambda^\nu_\rho p^\rho, \label{eq16} \\
\hat{p}'_\mu= \Lambda_\mu^\nu p_\nu
+\frac{1}{2\hbar}\bar{\theta}_{\mu\nu}\Lambda^\nu_\rho x^\rho
.\label{eq17}
\end{eqnarray}
 Inserting Eqs.(\ref{eq14}) and (\ref{eq15}) into Eqs.(\ref{eq16}) and (\ref{eq17}), one obtains
\begin{eqnarray}
\hat{x}'_\mu &=& \gamma (\Lambda_\mu^\nu \hat{x}_\nu
+\frac{1}{2\hbar}\Lambda^\mu_\nu\theta_{\nu\lambda}
\hat{p}^\lambda-\frac{1}{2\hbar}\theta_{\mu\nu}\Lambda^\nu_\lambda
\hat{p}^\lambda \nonumber\\
& + & \frac{1}{4\hbar^2}\theta_{\mu\nu}\bar{\theta}^{\lambda\alpha}\Lambda^{\nu}_\lambda \hat{x}_\alpha), \label{eq18} \\
\hat{p}'_\mu &=& \gamma (\Lambda_\mu^\nu \hat{p}_\nu
-\frac{1}{2\hbar}\Lambda^\mu_\nu\bar{\theta}_{\nu\lambda}
\hat{x}^\lambda-\frac{1}{2\hbar}\bar{\theta}_{\mu\nu}\Lambda^\nu_\lambda
\hat{x}^\lambda \nonumber\\
& - & \frac{1}{4\hbar^2}\bar{\theta}_{\mu\nu}
\theta^{\lambda\alpha}\Lambda^{\nu}_\lambda \hat{p}_\alpha).
\label{eq19}
\end{eqnarray}
Eqs.(\ref{eq18}) and (\ref{eq19}) define the non-commutative
Lorentz transformations on NC phase space, the
$\theta_{\mu\nu}p^\nu$ and $\bar{\theta}_{\mu\nu}x^\nu$ transform
as Lorentz vector on NC phase space. When $\bar{\theta}\rightarrow
0$, the Lorentz transformations (\ref{eq18}) and (\ref{eq19}) on
NC phase space return to the Lorentz transformations on NC space.
Using the inverse of the generalized Bopp's shift (\ref{eq14}) and
(\ref{eq15}), one finds that the square of the non-commutative
length on NC phase space is given by
\begin{equation}\label{eq20}
s^2_{ncps}=\gamma^2(\hat{x}^\mu \hat{x}_\mu
+\frac{1}{\hbar}\theta_{\mu\nu}\hat{x}^\mu \hat{p}^\nu
+\frac{1}{4\hbar^2}\theta^{\mu\alpha}\theta_{\mu\beta}\hat{p}_\alpha
\hat{p}^\beta)\;.
\end{equation}
One can check that $s^2_{ncps}$ is left invariant by the NC
Lorentz transformations on NC phase space.

It is straightforward to extend our results above to a Poincare
transformation, since a shift by a constant of the non-commutative
coordinates is compatible with the algebraic relations
Eq.(\ref{eq1}) on NC space
 and  Eq.(\ref{eq4}) on NC phase space. An infinitesimal non-commutative Poincare transformation
$\Lambda^\mu_\nu=\delta^\mu_\nu+\omega^\mu_\nu,
a^\mu=\epsilon^\mu$ is implemented by the operator
\begin{equation}\label{21}
U(1+\omega,\epsilon)=1+\frac{i}{2}\omega_{\mu\nu}J^{\mu\nu}-i\epsilon_\mu
p^\mu+\cdots
\end{equation}
with $J_{\mu\nu}=x_\mu p_\nu-x_\nu p_\mu$. The operator is
un-deformed, because the NC Poincare transformation is induced
  by the Poincare transformation on commuting space. So
the Lie algebra of the Lorentz group is also un-deformed,
\begin{eqnarray}\label{eq22}
&& [J_{\mu\nu},J_{\rho\sigma}] =
-i(\eta_{\nu\rho}J_{\mu\sigma}-\eta_{\mu\rho}J_{\nu\sigma}-
\eta_{\sigma\mu}J_{\rho\nu}+\eta_{\sigma\nu}J_{\rho\mu}) \; ,
\nonumber\\ && [p_\mu,J_{\rho\sigma}] = -i (\eta_{\mu\rho}
p_\sigma
- \eta_{\mu\sigma} p_\rho) \; , \nonumber\\
&& [ p_\mu , p_\nu] = 0 \;.
\end{eqnarray}

One may  apply our formulation to field theory. To do so,
derivatives have to be defined, by Ref.\cite{10}, the derivative
on NC space can be defined as
\begin{eqnarray}\label{eq25}
\hat{\partial}^x_\mu f(x,p) &=&
-i\theta^{-1}_{\mu\nu}[\hat{x}^\nu, f(x,p)]\nonumber
\\ &=& -i\theta^{-1}_{\mu\nu}[x^\nu-\frac{1}{2\hbar}\theta^{\nu\alpha}p_\alpha,
f(x,p)]\;.
\end{eqnarray}
Similarly, on NC phase space we  define the derivative of momentum
 as follows
\begin{eqnarray}\label{eq26}
\hat{\partial}^p_\mu f(x,p) & = &
-i\bar{\theta}^{-1}_{\mu\nu}[\hat{p}^\nu, f(x,p)] \nonumber\\
&=&
-i\bar{\theta}^{-1}_{\mu\nu}[[p^\nu+\frac{1}{2\hbar}\bar{\theta}^{\nu\alpha}x_\alpha,
f(x,p)]\;.
\end{eqnarray}
Under NC Lorentz transformations, the derivatives transform as
\begin{eqnarray}\label{eq27}
\hat{\partial}^{'x }_\mu &=&
\theta^{-1}_{\mu\alpha}\Lambda^\alpha_\beta\theta^{\beta\sigma}
\hat{\partial}^x_\sigma \; ,\nonumber\\
\hat{\partial}^{'p}_\mu &=&
\bar{\theta}^{-1}_{\mu\alpha}\Lambda^\alpha_\beta\bar{\theta}^{\beta\sigma}
\hat{\partial}^p_\sigma \;.
\end{eqnarray}

Now let's consider a non-commutative action for a Dirac fermion
coupled to a Yang-Mills gauge field, the action, in $2+1$
dimension, is given by
\begin{equation}\label{eq29}
 S = \int d^3
 \bar{\hat{\Psi}}(\hat{x})(i\hat{\! \not \! \! {D}}-m)\hat{\Psi}(\hat{x})
 - \frac{1}{2}\int
 d^3x Tr\hat{ F}_{\mu\nu}(\hat{x})\hat{F}^{\mu\nu}(\hat{x})\;.
\end{equation}
 Under the NC Lorentz
transformations, Ref.\cite{6} found that the NC Yang-Mills
potential transforms as
\begin{equation}\label{eq30}
\hat{A}^{'}_\mu=
\theta^{-1}_{\mu\alpha}\Lambda^\alpha_\beta\theta^{\beta\sigma}\hat{A}_\sigma,
\end{equation}
the NC covariant derivative transforms as
\begin{equation}\label{eq31}
\hat{D}^{'}_{\mu}= \theta^{-1}_{\mu\alpha}\Lambda^\alpha_\sigma
\theta^{\sigma\alpha}\hat{D}_\sigma,
\end{equation}
and  the field strength tensor  transforms as
\begin{equation}\label{eq32}
\hat{F}^{'}_{\mu\nu}=
\theta^{-1}_{\mu\rho}\Lambda^\rho_\sigma\theta^{\sigma\alpha}\theta^{-1}_{\nu\zeta}\Lambda^\zeta_\eta\theta^{\eta\beta}\hat{F}_{\alpha\beta}.
\end{equation}
as well as the NC spinor field transforms as
\begin{equation}\label{eq33}
\hat{
\Psi}^{'}=\exp(-\frac{i}{2}\omega^{\alpha\beta}S_{\alpha\beta})\hat{
\Psi},
\end{equation}
with $S_{\mu\nu}=\frac{i}{4}[\gamma_\mu,\gamma_\nu]$. If the
fields are taken in the enveloping algebra, the classical field,
namely, the leading order of Seiberg-Witten map, also transforms
according to (\ref{eq30})-(\ref{eq33}).  Though the equations
(\ref{eq30})-(\ref{eq33}) have the same  form  both for  $2+1$
dimensional NC space time and the $3+1$ dimensional NC space time
in \cite{6}, because of the singularity of the matrix
$\Theta=(\theta_{ij})$ in $3+1$ dimensional NC space-time , the
equations (\ref{eq30})-(\ref{eq33}) can not be applicable to $3+1$
dimensional NC space time, detailed discussions will be given in
the conclusion part.

To replace the noncommutative argument of the function by a
commutative one, we need to introduce a star product,
$f(\hat{x})g(\hat{x})=f(x)\star g(x)$, the star product here is
given in Eq.(\ref{eq2}) for NC space and defined in Eq.(\ref{eq8})
for NC phase space. It is easy to verify that these star product
is invariant under NC Lorentz transformations.
  After replacing NC coordinates  by the commutative ones through star product and expanding
the fields in the enveloping algebra by using Serberg-Witten Map,
the Eq.(\ref{eq29}), to the first order of $\theta$ ,  reads
\begin{eqnarray} \label{action}
S &=& \int d^3 x \bigg[ \bar \psi (i \! \not \! \! {D} - m )\psi -
\frac 1 4 \, \theta^{\mu \nu} \bar \psi F_{\mu \nu} (i \not \! \!
{D} - m )\psi \nonumber\\
& - & \frac{1}{2} \, \theta^{\mu \nu} \bar \psi \gamma^\rho
F_{\rho \mu} \, i {D}_\nu \psi
 -\frac{1}{2} \, {\rm Tr} \, F_{\mu \nu} F^{\mu \nu} \nonumber\\
&+& \frac{1}{4}  \, \theta^{\mu \nu} \, {\rm Tr} \, F_{\mu \nu}
F_{\rho \sigma} F^{ \rho \sigma} -  \, \theta^{\mu \nu} \, {\rm
Tr} \, F_{\mu \rho} F_{\nu \sigma} F^{ \rho \sigma} \bigg] + {\cal
O}(\theta^2)\; ,\nonumber\\
\end{eqnarray}
which is invariant under noncommutative Lorentz transformation.

\section{Conclusion remarks}

In this paper, we give a description of NC phase space, where the
star product can be easily defined( see Eq.(\ref{eq8})), the
Planck constant is modified on NC phase space( see
Eq.(\ref{eq5})).\\

 Following the Ref.\cite{6}, we define a non-commutative
Lorentz transformations both on NC space and NC phase space in
$2+1$ dimensions. The basic idea is to define the Lorentz
transformations for commutative coordinate (\ref{eq10}) and
momentum (\ref{eq11}) then to feed back these transformation to
the non-commutative sectors via variables transformations
(\ref{eq14}) and (\ref{eq15}). The algebraic relations both on NC
space and NC phase space are invariant under NC Lorentz
transformations, and the $\theta$-expanded gauge field action is
also invariant under NC Lorentz transformations.\\

 In this paper though the Greek indices are introduced,
we assume $\mu,\nu$ run as $i,j$, take values from 1 to 2, because
we set $\theta_{0\mu}=0$. Why we consider a 2-dimensional NC space
and NC phase space instead of 3-dimensional case. The main reason
is that on 3-dimensional NC space and NC phase space the
non-commutative parameters $\theta_{ij}$ and $\bar{\theta}_{ij}$
can be considered as anti-symmetric matrices, $\Theta
=(\theta_{ij}),\bar{\Theta} =(\bar{\theta}_{ij})$, the elements
can be written as
\begin{eqnarray}\label{34}
\theta_{ij}=\epsilon_{ijk}\theta_k,~~~\bar{\theta}_{ij}=\epsilon_{ijk}\bar{\theta}_k.
\end{eqnarray}
 i.e.
\begin{equation}\label{31}
\Theta
=\left(\begin{array}{ccc}0&\theta_3&-\theta_2\\
-\theta_3&0&\theta_1\\
\theta_2&-\theta_1&0
\end{array}\right)
,~~~\bar{\Theta}=\left(\begin{array}{ccc}0&\bar{\theta}_3&-\bar{\theta}_2\\
-\bar{\theta}_3&0&\bar{\theta}_1\\
\bar{\theta}_2&-\bar{\theta}_1&0
\end{array}\right).
\end{equation}
It is easy to find that these two anti-symmetric matrices are
singular matrices, because ${\rm det} \Theta ={\rm det}
\bar{\Theta} =0$. For these reason we can not define the inverse
of these matrices, and terms related to $\theta^{-1}_{ij}$ or
$\bar{\theta}^{-1}_{ij}$ will make no sense on 3-dimensional NC
space and  NC phase space. This problem  also exists in
Ref.\cite{6}, for example, the Eqs.[17-21] of the Ref.\cite{6}
make little sense in $3+1$ dimensions. The NC phase space
formulation and the NC Lorentz transformation of this paper can be
easily extended to any even dimensional space, however, for any
odd dimensional case, the anti-symmetric matrices $\Theta$ and
$\bar{\Theta}$ are singular, the method here would not be
applicable, and we will discuss it in our forthcoming studies.

\begin{acknowledgments}
Kang Li would like to thank Prof. S. Randjbar-Daemi for his kind
invitation and warm hospitality during his visit at the ICTP. This
work is supported in part by the National Natural Science
Foundation of China (10575026 and 10465004, 10665001). The authors
also grateful to the support from the Abdus Salam ICTP, Trieste,
Italy.
\end{acknowledgments}


\begin{thebibliography}{99}

\bibitem{1}N. Seiberg and E. Witten, JHEP {\bf 032 }, 9909(1999), hep-th/9908142.
\bibitem{2}M. Chaichian, M. M. Sheikh--Jabbari and A. Tureanu,
          Phys. Rev. Lett. {\bf 86 }, 2716(2001);  M. Chaichian, A. Demichev, P.
          Presnajder, M. M Sheikh-Jabbari and A. Tureanu, Nucl. Phys.
          B {\bf 611 }, 383(2001), hep-th/0101209.
\bibitem{3}M. Chaichian, P.
          Presnajder, M. M Sheikh-Jabbari and A. Tureanu, Phys. Lett.
          B {\bf 527}, 149 (2002) 149-154; H. Falomir, J. Gamboa, M. Loewe, F. Mendez and J. C. Rojas,
          Phys. Rev. D{\bf 66 },045018(2002), hep-th/0203260; Omer F. Dayi and Ahmed
          Jellal, J.Math.Phys. {\bf 43 },4592 (2002),
          hep-th/0111267.
\bibitem{4} Kang Li and Sayipjamal Dulat,{\it Eur. Phys. J.} C {\bf 46 } 825
(2006); Kang Li and Jianhua Wang, {\it Eur. Phys. J.} C {\bf 50 };
(2007)1007; {\it J. Phys. A: Math Theor.} {\bf 40 } (2007)2197.

\bibitem{5} B.Mirza and M.Zarei,   {\it Eur. Phys. J.} C {\bf 32 }
(2004)583; B.Mirza, R.Narimani and M.Zarei,{\it Eur. Phys. J.} C
{\bf 48 }(2006)641.

\bibitem{7} Kang Li, Jianhua Wang and Chiyi Chen, Mod. Phys. Lett. A
 {\bf 20},  2165-2174 (2005).
\bibitem{8} Michal Demetrian and Denis Kochan, hep-th/0102050; Jian-zu Zhang, Phys. Lett. B{\bf 584}, 204(2004).


\bibitem{6} X. Calmet, {\it Phys.Rev. D}
{\bf 71 } (2005) 085012.

\bibitem {wess}J. Wess, hep-th/0408080; M. Chaichian, P. P.
Kulish, K. Nishijima, and A. Tureanu, Phys. Lett. B 604, 98(2004).

\bibitem{luki}J. Lukierski, A. Nowicki, and H. Ruegg, Phys. Lett.
B293, 344(1992).



\bibitem{gomis} J. Gomis, T. Mehen, Nucl. Phys.B {\bf 591},
265(2000).

\bibitem{alvarez} L. Alvarez Gaume, J. L. F. Barbon, R. Zwicky,
JHEP {\bf 05}, 057(2001).

\bibitem{seiberg} N. Seiberg, L. Susskind, N. Toumbas, JHEP {\bf
06},044(2000).

\bibitem{10} L.Alvarez-Gaume and Spenta R. Wadia, {\it Phys. Lett. } {\bf B501 }
(2001)319;  S-C. Chu,  Nucl. Phys. B {\bf 550 }, 151 (1999).


\end{thebibliography}
\end{document}